\documentclass[aps, pra,twocolumn,reprint,superscriptaddress, nofootinbib]{revtex4-1}

\usepackage{graphicx}
\usepackage{siunitx}
\usepackage[dvipsnames]{xcolor}
\usepackage{verbatim}
\usepackage{float}
\begin{document}



\title{Standalone mobile quantum memory system}

\author{Martin Jutisz}
\email[]{martin.jutisz@physik.hu-berlin.de}

\affiliation{Institut für Physik and IRIS Adlershof, Humboldt-Universität zu Berlin, Newtonstr. 15, Berlin 12489, Germany}

\author{Alexander Erl}
\affiliation{Deutsches Zentrum für Luft- und Raumfahrt e.V. (DLR), Rutherfordstraße 2, 12489 Berlin, Germany}
\affiliation{Technische Universität Berlin, Institute for Optics and Atomic Physics, Hardenbergstraße 36, 10623 Berlin, Germany}


\author{Janik Wolters}
\affiliation{Deutsches Zentrum für Luft- und Raumfahrt e.V. (DLR), Rutherfordstraße 2, 12489 Berlin, Germany}
\affiliation{Technische Universität Berlin, Institute for Optics and Atomic Physics, Hardenbergstraße 36, 10623 Berlin, Germany}
\affiliation{AQLS – Advanced Quantum Light Sources, Guerickestr. 12, 10587 Berlin, Germany}

\author{Mustafa Gündoğan}
\email[]{mustafa.guendogan@physik.hu-berlin.de}
\affiliation{Institut für Physik and IRIS Adlershof, Humboldt-Universität zu Berlin, Newtonstr. 15, Berlin 12489, Germany}

\author{Markus Krutzik}
\affiliation{Institut für Physik and IRIS Adlershof, Humboldt-Universität zu Berlin, Newtonstr. 15, Berlin 12489, Germany}
\affiliation{Ferdinand-Braun-Institut (FBH), Gustav-Kirchoff-Str. 4, Berlin 12489, Germany}

\date{\today}

\begin{abstract}
We present the implementation and performance analysis of a portable, rack-mounted standalone warm vapor quantum memory system, that also includes the laser package, control electronics and data processing hardware. The optical memory is based on long-lived hyperfine ground states of Cesium which are connected to an excited state via the $D_1$ line at \SI{895}{nm} in a $\Lambda$-configuration.  The memory is operated with weak coherent pulses containing on average $<1$ photons per pulse. The long-term stability of the memory efficiency and storage fidelity is demonstrated at the single-photon level together with operation in a non-laboratory environment.
\end{abstract}

\maketitle

\section{Introduction}
Quantum memories (QMs)~\cite{Lvovsky2009, Lei2023} are central to many applications in quantum information science from quantum sensing~\cite{Mazelanik2021, Barzel2024}, computation~\cite{Nunn2013, Gouzien2021} to long-distance quantum networking~\cite{Sangouard2011, Muralidharan2016, Borregaard2020}. To date, many realizations of QMs have been demonstrated with different physical systems, ranging from cold atomic ensembles~\cite{Yang2016, Distante2017, Vernaz-Gris2018, Wang2019, Zhang2024} to solid-state systems~\cite{Gundogan2015, Jobez2015, Stas2022, Ortu2022, Zhou2023}. As a necessary element of quantum repeaters, these devices should be able to operate in non-laboratory environments, and as such their future deployment in space could advance global quantum communication networks~\cite{Gundogan2021, Liorni2021, Gundogan2021b, Wallnofer2022, Gundogan2024}.  In this context, warm-vapor QMs~\cite{Reim2011, Wolters2017, Kaczmarek2018, Finkelstein2018, Guo2019, Wang2022, Buser2022, Thomas2023, Esguerra2023} are particularly promising due to their low complexity and low size, weight and power (SWaP). Unlike many other systems, they do not require laser or cryogenic cooling, which makes them attractive for practical applications.
The key performance metrics for quantum memories include efficiency, fidelity, storage time, and multimode capacity. While there have been significant advancements in enhancing each of these metrics, here we shift the focus to another crucial aspect: the long-term stability and reliable operation of a fully integrated system outside a controlled laboratory setting. Although a rack-mounted memory system has been recently demonstrated~\cite{Wang2022}, to the best of our knowledge, operation of such a device with all of its subsystems fully integrated into a single, portable platform outside the laboratory environment has not been reported in the literature before. 
 
\section{System Design}
\begin{figure*}
\includegraphics[width=1.0\linewidth]{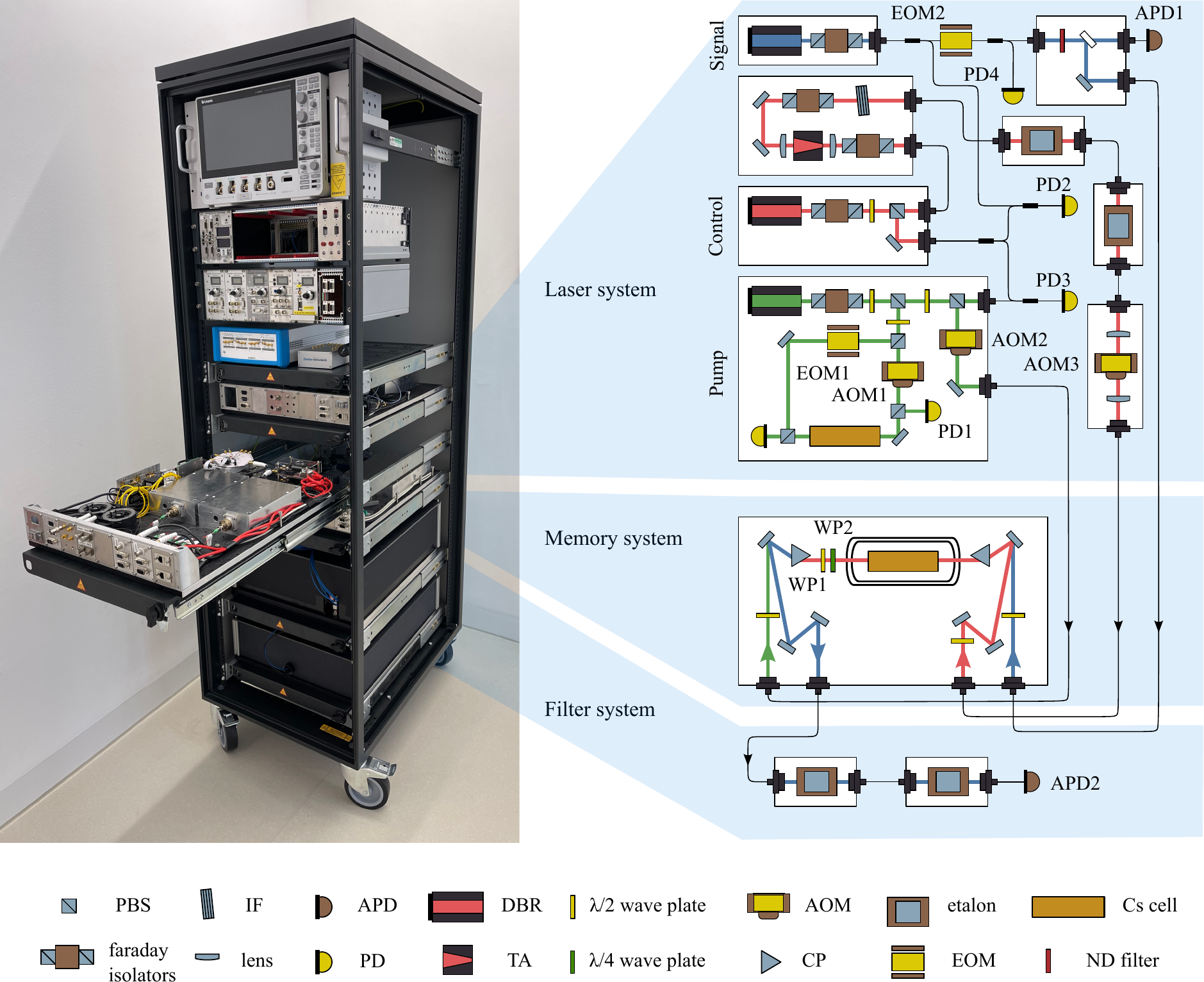}%
\caption{\label{Schematics}Schematics of system design including the laser system, the memory system and the spectral filtering system. PBS polarizing beam splitter; IF interference filter; APD avalanche photo diode; DBR distributed Bragg reflector laser; AOM acousto-optic modulator; PD photo diode; TA tapered amplifier; CP calcite prism; EOM electro-optic modulator; ND neutral density.}
\end{figure*}

The rack-mounted memory system, shown in figure \ref{Schematics}, features a laser system, a memory module, a spectral filtering module, and control electronics. The laser and the filtering systems are implemented with an emphasis on achieving high mechanical stability and robustness against environmental changes using similar technologies as reported in \cite{Pahl2019}. The control electronics include a fast arbitrary waveform generator, laser current driver and locking electronics, temperature controller, and a time tagger. The system is designed to function as a testbed for future implementation of miniaturized components with a reduced SWaP budget; therefore, a fast oscilloscope is included in the rack and can be used to optimize the memory operation with bright pulses for the single-photon level storage experiments. A custom-made, python-based software controls all aspects of the experiment. The individual optical subsystems are presented in the following sections.

\subsection{Laser system}
\begin{figure}[h]
\includegraphics[]{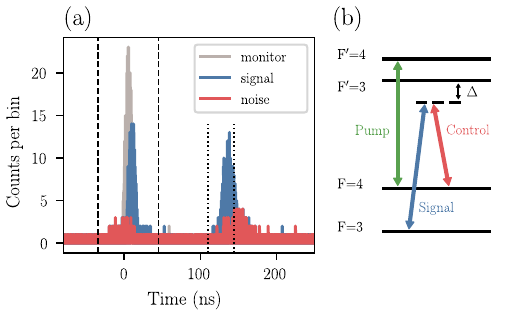}%
\caption{\label{histogram}
(a) Typical photon arrival histogram for a mean input photon number of $\mu_\text{in}=1.0(1)$ and a bin width of \SI{5}{ps}. The experimental parameters are listed in Table \ref{parameter}. The dotted lines represent the detection window used to calculate the efficiency and SNR of the retrieved signal pulses. The dashed line represents the detection window of the monitor pulse, used to determine the the input photon number $\mu_{in}$ (b) The hyperfine transition of the $^{133}$Cs $D_1$ line.}
\end{figure}

The compact laser system is housed in the central section of the rack depicted in figure \ref{Schematics} and includes three distributed Bragg reflector (DBR) laser diodes to provide the required optical fields at different frequencies for the pump, control and signal pulses. The pump laser serves as an absolute frequency reference, stabilized to the $F=4 \rightarrow F^{'} = 4$ hyperfine transition of the $^{133}$Cs $\text{D}_1$ line using frequency modulation spectroscopy detected on photodiode 1 (cf. figure \ref{histogram} (b)). The light is modulated using a free-space electro-optic modulator (EOM1) with a modulation frequency of \SI{7.4}{MHz}. An acousto-optic modulator (AOM2) driven at \SI{90}{MHz} is used to switch the pump laser, providing \SI{30}{mW} of optical power to the memory module. To ensure that the pump light is on resonance after the AOM2, a \SI{180}{MHz} AOM (AOM1) is included in the locking scheme as depicted in figure \ref{Schematics}. 

The control laser is offset-locked to the pump laser via the photo diode PD3 and amplified using a tapered amplifier (TA). In order to remove the  amplified spontaneous emission from the TA an interference filter (IF) with a full-width at half-maximum (FWHM) bandwidth of \SI{0.51(1)}{nm} and two monolithic Fabry-Pérot etalon with a finesse of $\mathcal{F}=50(5)$ and free spectral ranges (FSR) of \SI{51.6}{GHz} and \SI{25.8}{GHz} are used. Gaussian control pulses with a minimum FWHM of \SI{25}{ns} are temporally shaped using a \SI{200}{MHz} free-space AOM (AOM3). The maximum continuous wave optical power distributed to the memory module is \SI{70}{mW} and decreases with shorter pulses due to the reduction of the AOM efficiency.  

The signal laser is offset-locked to the control laser via PD2 and pulses are shaped using a fiber-coupled amplitude EOM (EOM2). The bias voltage of the EOM is locked to the minimum of the EOM transfer curve using a lock-in amplifier \cite{Snoddy2007} via PD4. Automated locking of all three lasers as well as the EOM bias is realized via a FPGA-based tool \cite{Wiegand2022}. A combination of neutral density (ND) filters are used to attenuate the bright input pulses to create weak coherent states (WCS) with controllable mean  input photon numbers, $\mu_\text{in}$, for storage experiments. A monitor path with a splitting ratio (memory path/monitor path) of $\theta = 11.2(4)$ enables the calibration of $\mu_\text{in}$ on an avalanche photo diode (APD1) with a detection efficiency of $\eta_{\text{APD, mon}}=36(5) \%$. 

\subsection{Memory system}
At the core of the memory system is a \SI{75}{mm}-long cylindrical glass cell with an inner diameter of \SI{4}{mm} and anti reflection coated windows  containing $^{133}$Cs atoms and \SI{5}{Torr} $\textrm{N}_2$ buffer gas. The cell is embedded in a copper rod and enclosed in three layers of mu-metal to prevent any spin wave decoherence through stray magnetic fields. Resistive heaters are mounted at the extremities of the rod that reach outside of the magnetic shield to allow heating of the cell without creating disruptive magnetic fields.

The perpendicular, linearly polarized, co-propagating signal and control beams are superposed via a calcite prism (CP) and focused at the center of the memory cell with a FWHM diameters of $\SI{120(5)}{\mu m}$ and $\SI{210(5)}{\mu m}$ respectively. Possible changes in the polarization due to the interaction with the atoms or the cell windows are counteracted using a combination of a quarter- and a half-wave plate (WP1 and WP2 in figure \ref{Schematics}). A collimated counter-propagating linearly polarized pump beam with a FWHM diameter of $\SI{678(5)}{\mu m}$ and a power of $\SI{30}{mW}$ overlaps with the signal and control beams.

\subsection{Filter system and detection}
To separate the strong control field from the signal field we use a combination of consecutive polarization and spectral filtering to reach a total suppression of the control field of about \SI{120}{dB}. Polarization filtering is achieved using a second calcite prism after the memory cell, resulting in a control beam suppression of $>\SI{60}{dB}$. For the subsequent spectral filtering we use two monolithic Fabry-Pérot etalons~\cite{Palittapongarnpim2012, Heller2023} with a free spectral range (FSR) of \SI{51.6}{GHz}, mounted in fiber-coupled housings. The etalon bandwidth is measured to be \SI{1.01(1)}{GHz} and \SI{1.16(2)}{GHz}. The transmission resonance is tuned by temperature to the signal frequency yielding a transmission of 90(1)\% and 87(1)\% respectively with a simultaneous suppression of the control frequency by about $\SI{30}{dB}$ per etalon. The temperature of the etalons is stabilized to $<\SI{0.4}{mK}$ and resonance shift is measured to be $<\SI{0.3}{MHz}$ over the course of several hours. For the design of the etalon mounts we focused on compactness and temperature stability in exchange for a non-ideal fiber coupling efficiency of 69(1)\% and 70(1)\%. We define the setup efficiency $\eta_{\text{setup}}=23(1)\%$ as the ratio between the power in the fiber connected to the APD (APD2) and the power right after the memory cell. The detection efficiency of the fiber-coupled APD2 is measured to be $\eta_{\text{APD, sig}}=30(5)\%$. 

\begin{table}[h]
\caption{\label{parameter}%
Experimental parameters used in the long duration operation.}
\begin{ruledtabular}
\begin{tabular}{ll}
Experimental parameter&Value\\
\hline
Signal pulse FWHM & \SI{10}{ns}\\
Control pulse FWHM & \SI{25}{ns}\\
Control peak power & \SI{5}{mW}\\
Delay of signal- with respect to control pulse & \SI{5}{ns}\\
Pump length & $\SI{30}{\mu s}$\\
Detuning (red) & \SI{1.5}{GHz} \\
Cell temperature & \SI{75}{\degreeCelsius}\\
Integration time $t_{\text{int}}$ & \SI{20}{s}\\
\end{tabular}
\end{ruledtabular}
\end{table}
\section{Memory optimization and characterization}

We start the experimental sequence by pumping the atoms in the $6^2 S_{1/2} F= 3$ ground state with a $\SI{30}{\mu s}$-long pulse resonant with the $F=4 \rightarrow F^{'} = 4$ transition and a power of \SI{30}{mW}. After waiting \SI{250}{ns} for the pump AOM to switch off completely, we send the write-in control pulse, centered at $t_0 =0$, to the memory. To ensure optimal temporal overlap between the signal and the write-in control pulse, the signal pulse is delayed with respect to the write-in control pulse by the time $t_d$. The FWHM of the Gaussian shaped signal pulses are set to $\tau_{\text{S}}^{\text{FWHM}} = \SI{10}{ns}$ to match the expected bandwidth of the SPDC photon source we want to couple to the memory in the near future \cite{Mottola2020}. A read-out control pulse is sent into the memory after a storage time $t_{\text{st}}$ to retrieve the stored excitation. A photon arrival histogram as shown in figure \ref{histogram} is recorded with an integration time of $t_{\text{int}}=\SI{20}{s}$, a bin width of \SI{5}{ps} and a sequence repetition rate, $f_{\text{rep}} \sim \SI{31}{kHz}$. We denote $N_{\text{mon}}$ as the sum over all counts of the monitor APD within a detection window of $\pm 4 \tau_{\text{S}}^{\text{FWHM}}$ around $t_0$, depicted as dashed line in figure \ref{histogram}. $N_{\text{sig}}$ is defined as the sum over all counts of the signal APD within a detection window of $t_{\text{st}} - 4 \tau_{\text{S}}^{\text{FWHM}}$ to $t_{\text{st}} -0.6 \ \tau_{\text{S}}^{\text{FWHM}}$ depicted as dotted line. We choose these limits as a good compromise between high efficiency and SNR. The noise counts $N_{\text{noi}}$ are measured with the same procedure but with the mean input photon number set to $\mu_\text{in}=0$.

\begin{figure*}
\includegraphics[]{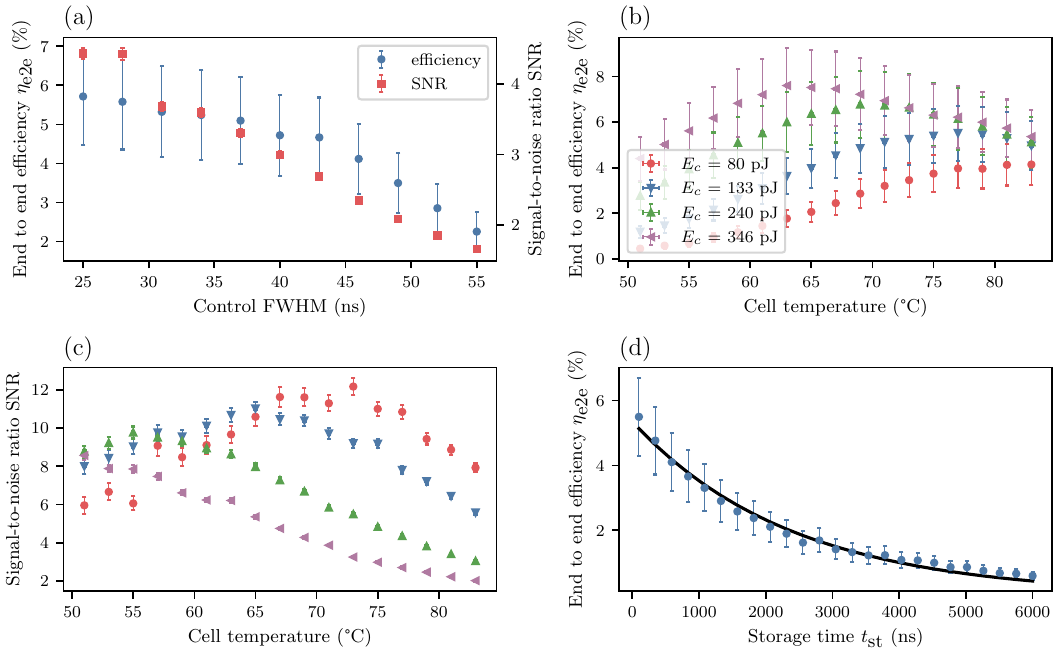}%
\caption{\label{memory_characterization}
Memory characterization for a mean input photon number of $\mu_\text{in}=1.2(2)$. Error bars include one standard deviation caused by the statistical uncertainty of the measurement and systematic uncertainties of the APD efficiencies and the ratio between the monitor and the memory path. The storage time is kept at $t_{\text{st}}=\SI{150}{ns}$ and timing between the signal and the control pulse $t_d$ is optimized for every measurement point of the panel (a), (b) and (d). (a) Efficiency as a function of the FWHM of the control pulses. The memory cell temperature is set to \SI{55}{\degreeCelsius}. The peak power of the Gaussian pulses is fixed at \SI{13}{mW}; (b) Memory efficiency and (c) SNR as a function of the memory cell temperature for different control pulse energies. The FWHM of the control pulse is set to \SI{25}{ns}; (d) Memory efficiency as a function of the storage time $t_{\text{st}}$. The black line represents an exponential fit with 1/e decay constants of \SI{2.4}{\mu s}. The experimental parameters are listed in Table \ref{parameter}.}
\end{figure*}

The mean input photon number $\mu_\text{in}= (N_{\text{mon}} \times \theta) / (t_{\text{int}} \times f_{\text{rep}} \times \eta_{\text{APD, mon}})$, the memory end-to-end efficiency $\eta_{\text{e2e}}=(N_{\text{sig}} - N_{\text{noi}})/(\mu_\text{in} \times \eta_{\text{APD, sig}} \times f_{\text{rep}} \times t_{\text{int}})$, and the signal-to-noise ratio $\text{SNR}=(N_{\text{sig}} - N_{\text{noi}})/ N_{\text{noi}}$ are then derived from the histogram. We note that the memory efficiency is slightly underestimated because the detection window for the input pulse is larger than that for the stored pulse. Additionally, the measured memory efficiency with WCS matches that with bright pulses, indicating the absence of input-intensity-dependent nonlinear effects.

To avoid fluorescent noise~\cite{Esguerra2023}, we choose to operate the memory outside the Doppler linewidth of the transition and set the detuning to $\Delta =$ \SI{1.5}{GHz}, a value that was found to be close to optimal in previous experiments \cite{Esguerra2023}. Operation at larger detunings would require higher optical power of the control pulse to reach the same efficiency~\cite{Reim2011}. We use Gaussian-shaped control pulses with their FWHM, timing and peak power being the only tunable parameters. However, higher efficiencies could potentially be reached using one of the different techniques of finding the optimal control pulse shape, including iterative pulse shaping \cite{Novikova2007}, machine learning \cite{Robertson2024, Lei2024} and the calculation of an optimal control field for a given signal pulse shape \cite{Phillips2008, Guo2019}. Figure \ref{memory_characterization} (a) shows the measured efficiency and \text{SNR} as a function of the control pulse's FWHM for a constant peak power of \SI{13}{mW} of the Gaussian control pulses at a cell temperature of \SI{55}{\degreeCelsius}. The best efficiency and \text{SNR} are reached for the shortest control pulses, suggesting that the speed of the pulse-shaping AOM might be limiting the memory efficiency and \text{SNR} at this cell temperature. We therefore set the control FWHM to the minimum value of $\tau_{\text{C}}^{\text{FWHM}} = \SI{25}{ns}$. The effective optical depth (OD) of the atomic ensemble can be either tuned by changing the cell temperature or the detuning from the transition. At higher temperatures, i.e. higher OD, no clear efficiency dependency on the control FWHM is observed within the measurement range. Note that the energy of the control pulses increases with increasing FWHM and constant peak power, resulting in larger control pulse energy and a larger noise contribution to the \text{SNR}.

For the theoretical case of optimal control pulses, the memory efficiency would only depend on the OD and the energy of the control pulse \cite{Gorshkov2007}. Figure \ref{memory_characterization} (b) \& (c) show the measured efficiency and the inferred fidelity at $\mu_\text{in}=1.2(2)$ as a function of the cell temperature for different control powers. Due to the correspondence between these two variables and the optimal temporal shape of the control pulse, the delay between the signal pulse and the control pulse $t_d$ is optimized for every measurement point. The memory efficiency peaks at a lower temperature for increasing optical power. At low OD the  memory efficiency seems to be limited by optical power of the control pulses. At higher ODs the memory efficiency decreases with increasing OD and saturates for increasing pulse energies. This behaviour might be explained by time-varying AC Stark shift \cite{Ming2023} at higher optical powers, re-absorption of the signal due to imperfect overlap of signal and control beams, increased collision rate between Cs atoms~\cite{Klein2009, Wang2022} or a non-optimal control pulse shape~\cite{Guo2019}. The SNR peaks at a lower OD and power compared to the efficiency caused by the fast increase of noise counts with increasing control power and OD. 
Noise contribution may include fluorescent noise, spontaneous Raman scattering  \cite{Esguerra2023} and four-wave-mixing noise \cite{Lauk2013, Thomas2019}. Improving the pumping efficiency by using a pump beam that covers the entire cell or optimizing the amount of buffer gas is expected to reduce the noise. Four-wave-mixing noise could be reduced by exploiting polarization selection rules \cite{Buser2022}.

\begin{figure}[t]
\vspace{20pt}
\includegraphics[]{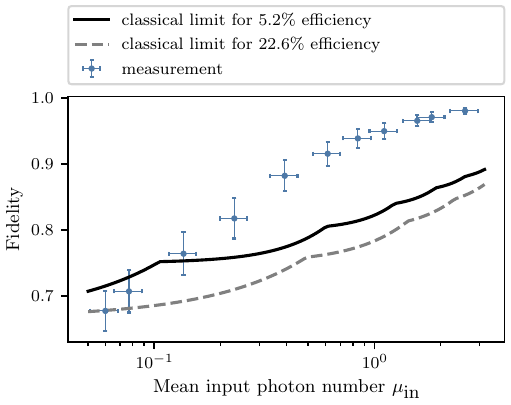}%
\caption{\label{fidelity} Inferred fidelity as a function of the mean input photon number for a storage time of $t_{\text{st}}=\SI{150}{ns}$, the black line represents the fidelity limit of a classical strategy for the measured end-to-end efficiency of $\eta_{\text{e2e}}=5.2(1.1)\%$. The dashed line shows the equivalent limit for the corresponding memory efficiency of $\eta_{\text{mem}}=22.6(5.3)\%$. Error bars include one standard deviation 
caused by the statistical uncertainty of the measurement and systematic uncertainties of the APD efficiencies and the ratio
between the monitor and the memory path.}
\end{figure}

For the following measurements we chose the experimental parameters summarized in Table \ref{parameter} to simultaneously reach a high efficiency and SNR. Figure \ref{memory_characterization} (d) shows the memory efficiency as a function of the storage time. We fit the data with an exponential function $f(t_{\text{st}})=H e^{-t_{\text{st}} \tau}$ to extract the memory lifetime $\tau = \SI{2.4(1)}{\mu s}$ and the zero time efficiency of $H_{\text{e2e}}=5.4(1)\%$, corresponding to a memory efficiency of $H_{\text{mem}}= H_{\text{e2e}} / \eta_{\text{setup}} = 23(1)\%$. The errors correspond to one standard deviation of the fit function. We assume the lifetime to be limited by the diffusion of the atoms out of the interaction zone which could be increased by using larger beams or a higher amount of buffer gas.

To test the quantum character of the memory, we infer the fidelity of the storage process from the achieved signal-to-noise ratio (SNR) similar to Refs.~\cite{Wang2022, Thomas2023}. The fidelity in this case is given by $F = (\mu_{\text{in}} + \mu_1)/(\mu_{\text{in}} + 2\mu_1)$, where $\mu_1 = N_{\text{noi}} / (\eta_{\text{e2e}} \times f_{\text{rep}} \times t_{\text{int}})$ is the mean photon number at the input that would yield a SNR of 1 at the output~\cite{Gundogan2015, Jobez2015}. We then benchmark this value against a modified fidelity threshold for WCS~\cite{Specht2011, Gundogan2012} which also takes into account the finite memory and transmission efficiencies. Figure ~\ref{fidelity} shows that the inferred fidelity in our experiments surpasses this threshold beyond $\mu_{\text{in}}>0.2$ when the transmission efficiency is included in the calculation. The error bars are calculated in the same way as for the previous measurements. This demonstrates that the performance of the memory cannot be mimicked by a classical measure and prepare strategy~\cite{Massar1995} as it would inherently add excess noise~\cite{Peres2003} and therefore preclude~\cite{Dieks1982, Wootters1982} any high-fidelity operation.  We also note that no background subtraction~\cite{Leung2024} has been applied to any of these results.

\begin{figure}[]
\includegraphics[]{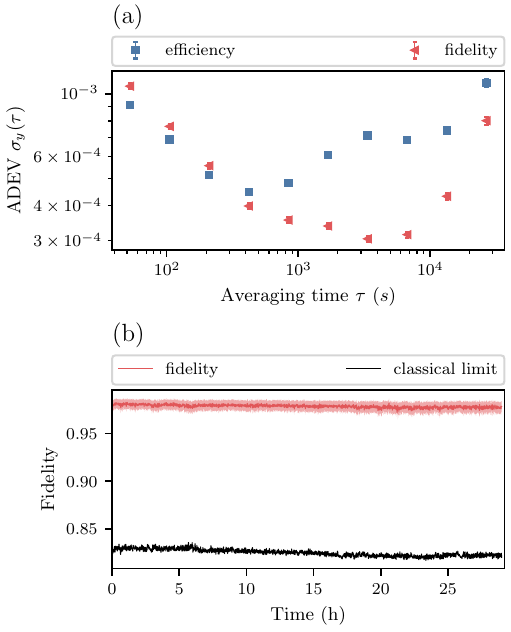}%
\caption{\label{ADEV} Long-term memory stability outside of a laboratory environment. (a) Overlapping Allan deviation of the end-to-end efficiency $\eta_{\text{e2e}}$ and the fidelity. The error bars are calculated using Chi-squared statistics. (b) Time trace of the fidelity with the fidelity limit of a classical device. The shaded region represents the confidence interval including one standard deviation caused by the statistical uncertainty of the measurement and
systematic uncertainties of the APD efficiencies and the ratio
between the monitor and the memory path. }
\end{figure}

\section{Long duration operation in non- laboratory environment}
To benchmark the robustness of the system under uncontrolled environmental conditions we moved the rack to an empty office room in the same building without any temperature and humidity control. We demonstrate the stability of the memory system by repeatedly measuring every \SI{53.1(2)}{s} the efficiency and fidelity over the duration of \SI{28}{hours}. Due to an unfortunate coincidence, the measurement took place during a time 
of high outside temperatures ($>\SI{30}{\degreeCelsius}$) which lead to an elevated room temperature up to \SI{34.1(1)}{\degreeCelsius} which corresponds to a temperature difference of $>\SI{10}{\degreeCelsius}$ compared to the laboratory temperature. Nevertheless, memory operation could be resumed and maintained without the need for realignment. 
Over the duration of this measurement the average storage efficiency of $\eta_{\text{e2e}}=5.1(1.1)\%$ ($\eta_{\text{mem}}=22.1(5.2)\%$) and the average input photon number of $\mu_\text{in}=1.0(1)$ matched the laboratory performance. The wave plates (WP1 and WP2) before the polarization filtering were optimized for the specific cell operation temperature of $\SI{75}{\degreeCelsius}$ by minimizing the noise count rate on the APD2 improving the average SNR to $14.4(5)$ corresponding to a noise level of $\mu_1 = 0.023(5)$ and a average fidelity of $F = 0.98(1)$ . 

To investigate the memory stability performance we calculate the overlapping Allan deviation (ADEV) \cite{riehle2006} of the memory end-to-end-efficiency and inferred fidelity shown in figure ~\ref{ADEV} (a) out of the recorded time trace. The instability of the efficiency reaches its minimum at an averaging time of $\tau = \SI{1}{h}$ and stays below $\sigma_y(\tau) < 1.1 \times 10^{-3}$ over the measurement range. For longer averaging times the instability of the efficiency and fidelity increases due to thermal fluctuations mainly causing polarization fluctuations in the optical fibers that induce power fluctuations at the memory input. \textcolor{black}{figure ~\ref{ADEV} (b) shows the long-term stability of the inferred fidelity and the corresponding classical threshold. The fidelity stays mostly constant at around $98\%$ during the 28 hour-long measurement. The fluctuations of the classical threshold are due to small variations of the mean input photon number and the storage efficiency.}

\section{Conclusion and outlook}
We have developed a portable, rack-mounted, standalone warm vapor QM system and demonstrated its stable operation outside of a controlled laboratory environment. This represents a notable step forward, enabling quantum memory experiments in a variety of settings without the need for specialized laboratory conditions. The system’s long-term stability in non-controlled environments indicates its potential for practical applications in quantum communication. In light of recent long-range entanglement experiments~\cite{Rakonjac2023, Arenskotter2023, Liu2024, Knaut2024} involving QMs, this transportable platform could enable experiments in settings without the need for access to a high-quality research laboratory. Additionally, recent advances~\cite{Thomas2023, Thomas2024, Maass2024} towards storing single photons emitted by quantum dot sources in warm vapor QMs when combined with multimode storage capability~\cite{Messner2023} would  enable the realization of enhanced quantum repeater schemes~\cite{Sangouard2007, vanLoock2020}.

Future work will focus on miniaturizing the system~\cite{ Christ2024} to reduce its size for deployment on small-scale satellite platforms~\cite{Oi2017b}. This would allow the system to be used in space-based quantum networks. In parallel, incorporating a miniaturized photon-pair source \cite{Villar2020, Mottola2020} could enable the system to function as a fully self-contained unit. Efforts will also be made to further improve storage efficiency, fidelity, and control pulse shaping~\cite{Robertson2024, Lei2024}, as well as reduce noise~\cite{Buser2022}, ensuring the system performs optimally under a range of conditions.

\section{ACKNOWLEDGEMENTS}
We would like to thank Lukas Heller for his support in the filter cavity design and Patrick Ledingham for discussions. We acknowledge the support from DLR through funds provided by BMWK (OPTIMO-III and QuMSeC), No.~50WM2347 and No.~50RP2090, funding by state of Berlin (application number 10206872), funding by the state of Berlin within the Pro FIT program under grant number 10206829 and by the Federal Ministry of Education and Research (BMBF) under grant number 16KISQ040K. M.G. acknowledges funding from the European Union's Horizon 2020 research and innovation program under the Marie Skłodowska-Curie grant agreement No.~894590 (QSPACE) for the support during the early phases of this work.

\bibliography{qumsec}

\end{document}